\documentclass[journal]{IEEEtran}

\usepackage{amsmath}
\newtheorem{theorem}{Theorem}[section]

\newtheorem{lemma}[theorem]{Lemma}

\begin{document}

\title{TQL: Towards Type-Driven Data Discovery}

\author{Andrew Y. Kang, Sainyam Galhotra\thanks{Cornell University, e-mail: {ayk36@cornell.edu}, sg@cs.cornell.edu}}
\maketitle



\begin{abstract}
    Existing query languages for data discovery exhibit system-driven designs that emphasize database features and functionality over user needs. We propose a re-prioritization of the client through an introduction of a language-driven approach to data discovery systems that can leverage powerful results from programming languages research. In this paper, we describe TQL, a flexible and practical query language which incorporates a type-like system to encompass downstream transformation-context in its discovery queries. The syntax and semantics of TQL (including the underlying evaluation model), are formally defined, and a sketch of its implementation is also provided. Additionally, we provide comparisons to existing languages for data retrieval and data discovery to examine the advantages of TQL's expanded expressive power in real-life settings.
\end{abstract}




\section{Data Discovery: Limited By Language}
Effective data discovery~\cite{behme2024fainder} is critical to the success of modern data science. With the ever-increasing availability of data and the ever growing primacy of data-driven operations and decision-making across organizations, the methods and procedures for identifying and retrieving relevant high-quality data have now become critical bottlenecks to data analytics and modeling (e.g. causal inference and/or machine learning). Indeed, recent research has only confirmed that data discovery remains one of the most time-consuming and labor-intensive aspects of the data science pipeline \cite{Hulsebos}.

In most real-world scenarios, it is exceedingly rare that the data discovery user (data scientist, data analyst, etc.) has a concrete understanding of the datasets most appropriate to their particular problem~\cite{galhotra2023metam}. Rather, it is more often the case that the prior knowledge of the user-at-hand is limited to a set of identified properties and relationships which are expected to encapsulate their desired datasets. 

The role of the query language in a data discovery system is therefore to provide a method through which the user's pre-existing knowledge might be communicated and applied to best guide the semi-automated discovery process. It should be apparent then that a well-designed query language should enable the user to express relevant properties and relationships with the greatest possible clarity, the greatest degree of precision, and the greatest expressive flexibility. 

The design of effective syntax and semantics to meet these priorities has been well-studied in traditional programming languages research \cite{Stefik,Felleisen}. Existing query languages, however, fail to leverage the results of such research. Data discovery systems typically consist of both a query language and an underlying model upon which it can be evaluated \cite{Paton}; and in most cases, this language-model complement is accomplished by the subordination of the former to the latter. For example, in the data discovery system Aurum, the provided SRQL language is constructed in correspondence to the layered enterprise knowledge graphs upon which the data is modeled. Similarly, the SPARQL language was explicitly designed to operate on resource description framework representations of data \cite{Perez}.

While model-driven designs for discovery systems can streamline the pathway to implementation and often do inspire innovations in data representation, such as in the case of the enterprise knowledge graphs of Aurum \cite{Fernandez}, it is nevertheless the case that this design methodology greatly limits the flexibility of the associated query language to the detriment of the general-purpose applicability of the discovery system. In particular, we highlight some consequences of interest from a programming languages perspective:

\begin{itemize}
    \item \textbf{Poor Language Design.} In model-driven designs, the syntax and semantics of the query language are effectively reduced to a basic correspondence between language and model elements. As a result, modern programming languages techniques are difficult to leverage to increase expressive power or to provide other quality-of-life features.
    
    \item \textbf{Lack of Standardization.} Since the underlying model for different data discovery systems are almost always necessarily distinct, the query languages of model-driven designs tend to share few consistent features and struggle to follow some consistent or common subset of paradigms. A user of data discovery systems is therefore compelled to relearn a new set of syntax, semantics, and idioms whenever they need to switch discovery systems to better meet the requirements of their task-at-hand.
\end{itemize}

Traditional programming languages (e.g. Java, Python, C) are organized into a few general classes of paradigms (e.g. procedural, functional, object-oriented) which tend to each be specialized towards particular software engineering goals \cite{Bartonicek}. Owing to the diversity of available data and representations, we expect data discovery systems and their associated query languages to form an equally diverse ecosystem. In this paper, we thus avoid any pretensions of constructing a universal query language, and instead content ourselves to presenting a data discovery and retrieval language for tabular/relational data designed in accordance to programming languages principles. 

To do so, we introduce TQL, a new powerful and practical query language which utilizes a type-system to encode downstream transformation-context into a discovery search space, thereby synthesizing features from across both data retrieval and data discovery languages. TQL is designed to be modular and extendable, and its syntax is designed to be intuitive to existing discovery system users, following SQL in its rough outline, but also taking limited inspiration from the functional programming language OCaml for its additional features.

\section{Type-Collected Relational Algebra}

In this section, we introduce type-collected relational algebra. Type-collected relational algebra is an extension of traditional relational algebra that describes the evaluation model for our Type-driven Query Language (TQL). 

In the subsequent subsections, we present type-collected relational algebra as a declarative evaluation model. We begin by providing the necessary justification and intuition for reasoning with the model. We then formally define the components of the model in mathematical notation. Using this formalization, we prove that TQL is a relationally-complete language. 

\subsection{Overview}

To better reason with type-collected relational algebra, we provide some justification for its inclusion and offer some intuition as to its structure.

For a language to properly encode transformation-context, it must be able to express the basic transformations corresponding to the type of queried data upon which it acts. It thus follows that the ability of a language for tabular data to encode transformation-context would follow by its capacity to express relational algebra operations.

The traditional construction of relational algebra, however, requires pre-existing knowledge of the datasets to be used --- the very information that, if known, would fulfill the purpose of the data discovery system and thus make it redundant. Hence the disconnect between search and transformation in traditional query and discovery languages. To bypass this segregation, a new approach is therefore necessary.

Type-collected relational algebra provides such an approach. Intuitively, the model can be understood as functioning like traditional relational algebra, but with operations instead acting upon dataset variables in the place of explicit dataset declarations. These variables, representing undetermined datasets in the query statement, are modeled as collections containing all potential datasets which might satisfy that variable.  Collection-level operations can thus be intuitively defined in terms of a corresponding dataset-level function. 

Additionally, each collection is "typed" to the extent that it has a corresponding predicate denoting the particular properties and relationships of its contained datasets. More specifically, predicates are enforced for collections in accordance to the constraint-based approach of the Aurum data discovery system \cite{Fernandez}. That is, tables in a collection must satisfy a series of selection criteria, of which there are two types: property constraints, concerning the particular characteristics of a table; and relationship constraints, which concern relationships between the table and some other fixed tables. 

\subsection{Notation \& Definitions}
We introduce the notation and formal definitions for the essential elements of type-collected relational algebra:

\begin{itemize}
\item Collections, which are sets of datasets.
\item Operations, which are functions that take in 1 or 2 collections as input, returning a single collection as output.
\item Predicts, which are applied to collections to enforce property and relationship constraints.
\end{itemize}

\noindent \textbf{Definition 1} (Collections). We formally define a collection $C = \{t_1, t_2, \dots, t_n \}$ as a set of some datasets (or tables) $t_1, t_2, \dots, t_n$. Given two collections $C_1$, and $C_2$, we permit collections to be combined in type-collected relational algebra by the normal set operations $C_1 \cup C_2$, $C_1 \cap C_2$, and $C_1 - C_2$. Notation-wise, we use capital letters to denote elements related to collections, and lowercase letters for those concerning tables. \\

\noindent \textbf{Definition 2} (Operations). Let $f(t_0)$ and $g(t_1, t_2)$ denote (possibly partial) functions on tables $t_i$. The corresponding operation on collections are formally defined as the functions: $f(C_0) = \{ f(t) : t \in C_0 \}$ and $g(C_1, C_2) = \{ g(t_1, t_2): t_i \in C_i \}$. \\

\noindent \textbf{Definition 3} (Predicates). Let $p(t)$ denote a predicate representing table $t$'s satisfaction of a series of constraints. Then, given a collection $C$, we say that $p(t)$ restricts $C$ to $C_{p(t)}$ under the following formal definition: $C_{p(t)} = \{ t \in C : p(t) \}$. The special notation $C_{[t]}$ for predicates denotes that collection $C$ has been restricted to only include table $t$.

\subsection{Expressive Power Guarantees}
In this section, we establish a baseline in expressive power by formally proving that type-collected relational algebra is relationally-complete. \\

\noindent \textbf{Equivalence}. To prove that type-collected relational algebra is relationally-complete, we must show that it can express a certain subset of relational algebra expressions \cite{Aho}. Expressions in traditional relational algebra, however, act upon individual datasets, whereas expressions in type-collected relational algebra act upon collections of datasets. As such, we must first establish a correspondence between tables and single-element collections. \\

\begin{lemma}
    Operations in type-collected relational algebra are closed on single-element collections for all tabular values for which they are defined. \\
\end{lemma} 

\noindent \textit{Proof}. By definition, each table in the output set of a collection operation is constructed by selecting a table from each input collection to serve as input tables for the corresponding table-level function. Thus, if each input collection consists of only one element, then the output collection consists of either one element (in which case, the lemma holds), or no elements (in which case, it by definition must be the case that the table-level function was not defined for those tabular values, and so the lemma holds as well). \\

\noindent \textbf{Relational Completeness}. To show that type-collected relational algebra is relationally-complete, we show that each of the expressions in traditional relational algebra has a corresponding expression in type-collected relational algebra. Completeness then follows by lemma on closure. \\

\begin{theorem} The following relational algebra expressions, which are necessary to show equivalence with relational calculus and thus relational completeness, are each expressible in type-collected relational algebra:
\begin{enumerate}
\item Cartesian Product: $p(t_0, t_1) := t_0 \times t_1$
\item Set Union: $u(t_0, t_1) := t_0 \cup t_1$
\item Set Difference: $d(t_0, t_1) := t_0 - t_1$
\item Selection: $\sigma_F (t_0)$
\item Projection: $\pi_{s_1, s_2, \dots, s_n} (t_0)$
\end{enumerate}
\end{theorem}

\noindent \textit{Proof}. Per the lemma, there is an equivalence between single-element collections in type-collected algebra and datasets in relational algebra. As such, given any dataset $t$, it is the case that $t \equiv C_{[t]}$.

Then, by the equivalence established in the lemma:
\begin{itemize}
\item $p (C_{[t_0]}, S_{[t_1]}) \equiv t_0 \times t_1$
\item $u (C_{[t_0]}, S_{[t_1]}) \equiv t_0 \cup t_1$
\item $d (C_{[t_0]}, S_{[t_1]}) \equiv t_0 - t_1$
\item $\sigma_F (C_{[t]}) \equiv \sigma_F(t)$
\item $\pi_{s_1, s_2, \dots, s_n} (C_{[t]}) \equiv \pi_{s_1, s_2, \dots, s_n}(t)$
\end{itemize}

\subsection{Type Inference for Predicates}
We defer a full discussion of type inference on predicates, for which research is still ongoing, to a future paper. Here, we provide a sketch of the theory motivating this endeavor. From the Curry-Howard Isomorphism \cite{Wadler}, it is known that there exists an intimate relationship between predicate logic and type systems. As such, it must follow that given a type system for which there are a set of inference rules, there should exist a matching method of inference for the predicates to which this type system corresponds.

Empirical evidence for such an application of the isomorphism is provided by existing interdisciplinary work between programming languages and computer security at the intersection of information flow control. JFlow, a security language extension for Java, for example, employs a system of type rules to infer complex predicates concerning information flow across a variety of data operations \cite{Myers}.

\section{Type-driven Query Language (TQL)}

\subsection{Syntax}
We first introduce the syntax for TQL in Backus-Naur form. 

\begin{equation} \notag
    \begin{split}
        Q ::=& \; C; \text{ $|$ } C; \; Q \\
        C ::=& \; id \text{ $|$ } fnc \text{ $|$ } id = C \text{ $|$ } C : \{ sig \} \\
        |& \; C_0 \texttt{ AND } C_1 \text{ $|$ } C_0 \texttt{ OR } C_1 \text{ $|$ } C_0 \texttt{ NAND } C_1 \\
        fnc ::=& \; \texttt{SELECT}[lst] \; C \text{ $|$ } \texttt{UNION } C_0 \; C_1 \text{ $|$ } \texttt{DIFF } C_0 \; C_1  \\
        |& \; \texttt{PROD } C_0 \; C_1 \text{ $|$ } \texttt{FILTER}[pd] \; C_0 \\
        |& \; \texttt{JOIN } C_0 \; C_1 \text{ $|$ } \texttt{JOIN}[pd] \; C_0 \; C_1 \\
        sig ::=& \; prp \text{ $|$ } \texttt{NOT } sig \text{ $|$ } sig_0 \texttt{ AND } sig_1 \text{ $|$ } sig_0 \texttt{ OR } sig_1 \\
        prp ::=& \; \texttt{SRC}[str] \text{ $|$ } \texttt{COL}[str] \text{ $|$ } \texttt{COL*}[str] \text{ $|$ } \texttt{SIML}[id] \\
        |& \; \texttt{PFKEY}[id] \text{ $|$ } \texttt{FORALL}[pd] \text{ $|$ } \texttt{EXISTS}[pd] \\
        pd ::=& \; e_0 \; cmp \; e_1 \text{ $|$ } \texttt{NOT } pd \text{ $|$ } pd_0 \texttt{ AND } pd_1 \text{ $|$ } pd_0 \texttt{ OR } pd_1 \\ 
    \end{split}
\end{equation}

\begin{equation} \notag
    \begin{split}
        cmp ::=& \; \geq \text{ $|$ } > \text{ $|$ } \leq \text{ $|$ } < \text{ $|$ } = \text{ $|$ } \neq \\  
        e ::=& \; val \text{ $|$ } id[str] \text{ $|$ } e_0 \oplus e_1 \\
        lst ::=& \; str \text{ $|$ } str, \; lst
    \end{split}
\end{equation}

\begin{equation} \notag
\begin{split}
id &\text{ \; collection identifier} \\
val &\text{ \; float, integer, or string} \\
str &\text{ \; string } \\
\end{split}
\end{equation}

\subsection{Semantics}
In TQL, programs are evaluated queries (denoted by $Q$ in Backus-Naur Form). We define the semantics of each syntactic construction introduced in the previous part from the top-down, referencing type-collected relational algebra wherever possible. Evaluation for more complicated compound queries then follows implicitly by structural induction.

\subsubsection{Queries} $\,$ \\
As defined in the Backus-Naur Form, a query $Q$ is either $C_0 \texttt{;}$, a collection $C_0$, or $C_1 \texttt{;} Q_1$, a collection $C_1$ followed by another query $Q$. In the former case, the program returns collection $C_0$ as its output; in the latter case, the program returns the result of $Q_1$, where $C_1$ may be referenced in $Q_1$ using the appropriate collection identifier.

\subsubsection{Collections} $\,$ \\
We describe the semantic evaluation for each possible construction of a collection $C$:
\begin{itemize}
\item $id$ is a reference to the collection whose name is identified by $id$. If $id$ has not already been previously assigned a collection, then it is initialized as follows in type-collected relational algebra: $id = S$, where $S$ refers to the collection of all datasets from all possible sources (as it was defined in the model).
\item $id = C$ assigns $id$ to reference $C$. In type-collected relational algebra, it corresponds to $id = C$. 
\item $C : \{ sig \}$ applies the restrictions of signature $s$ to the collection $C$. That is, it corresponds to $C_{sig}$ in type-collected relational algebra. 
\item $fnc$ is the collection returned by the operation $fnc$.
\item $C_0 \texttt{ AND } C_1$ corresponds to $C_0 \cap C_1$ in type-collected relational algebra.
\item $C_0 \texttt{ OR } C_1$ corresponds to $C_0 \cup C_1$ in type-collected relational algebra.
\item $C_0 \texttt{ AND NOT } C_1$ corresponds to $C_0 - C_1$ in type-collected relational algebra.
\end{itemize}

\subsubsection{Functions} $\,$ \\
Each function $F$ evaluates to a collection $C$. We define the evaluation of each function as follows:
\begin{itemize}
\item $\texttt{SELECT}[lst] \; C$ corresponds to $\pi_{a_1, a_2, \dots, a_n} (C)$ in type-collected relational algebra, where $a_1, a_2, \dots, a_n$ are the attributes (i.e. columns) listed in $lst$. 
\item $\texttt{FILTER}[pd] \; C$ corresponds to $\sigma_{pd} (C) $ in type-collected relational algebra, where $pd$ is a predicate on rows. 
\item $\texttt{UNION } C_0 \; C_1$ corresponds to the operation $u(C_1, C_2)$ in type-collected relational algebra. 
\item $\texttt{DIFF } C_0 \; C_1$ corresponds to the operation $d (C_1, C_2)$ in type-collected relational algebra. 
\item $\texttt{PROD } C_0 \; C_1$ corresponds to the operation $p(C_1, C_2) $ in type-collected relational algebra. 
\item $\texttt{JOIN} \; C_0 \; C_1$ and $\texttt{JOIN}[pd] \; C_0 \; C_1$ correspond to $j (C_1,C_2)$ and $j_{pd} (C_1,C_2)$, where $j(t_1, t_2)$ denotes a join operation on $t_1$ and $t_2$, parametrized by predicate $pd$. 
\end{itemize}

\subsubsection{Signatures} $\,$ \\
In TQL, a signature $sig$ applies a restriction to a collection $C$ by type-collected relational algebra search action, $C_{sig}$. Semantically, the syntactic construction for signatures in TQL describe predicates which can be used to apply that search action. They are constructed recursively as follows:
\begin{itemize}
\item A signature can be formed from a single constraint (or property): $prp$
\item A signature can be defined as the logical negation of another signature: $\texttt{NOT } sig$
\item A signature can be defined as the logical AND of two other signatures: $sig_0 \texttt{ AND } sig_1$
\item A signature can be defined as the logical OR of two other signatures: $sig_0 \texttt{ OR } sig_1$
\end{itemize}

For each table $t$ in a collection, we define the constraints $prp$ as follows:
\begin{itemize}
\item $\texttt{SRC}[str]$ evaluates to true if the table $t$ is the table named $str$, or if it originates from a table named $str$ (e.g. as the result of a join operation).
\item $\texttt{FORALL}[pd]$ evaluates to true if the predicate $pd$ holds for every row of table $t$.
\item $\texttt{EXISTS}[pd]$ evaluates to true if the predicate $pd$ holds for some row of table $t$.
\item $\texttt{COL}[str]$ and $\texttt{COL*}[str]$ respectively evaluate to true if there exists a column named $str$ or a column that contains the keyword $str$.
\item $\texttt{PFKEY}[id]$ evaluates to true if there exists some primary-foreign key pair upon which $t$ can be joined to a table $\tilde{t}$ in the collection $id$.
\item $\texttt{SIML}[id]$ evaluates to true if $t$ is sufficiently similar to some table $\tilde{t}$ in collection $C$ as determined by some implementation-dependent similarity measure (e.g. Jaccard Similarity).
\end{itemize}

\subsubsection{Row Predicates \& Expressions} $\,$ \\
A row predicate $pd$ is a statement which must evaluate to a truth value of true or false. It can be constructed as follows:
\begin{itemize}
\item $e_0 \; comp \; e_1$, where $comp$ is a comparison operator, is the atomic (or primitive) form of a row predicate. It evaluates to true if the evaluated values for $e_0$ and $e_1$ satisfy the comparison, and false otherwise. 
\item $\texttt{NOT } pd$, as well as $pd_0 \texttt{ AND } pd_1$, and $ pd_0 \texttt{ OR } pd_1$ are used to form compound predicates from other predicates, corresponding to the logical NOT, AND, and OR operations respectively.
\end{itemize}

We now define expressions. Expressions can be formed from values $v$, from attributes $a$, or from the application of arithmetic binary operators to two seperate expressions | that is: $e_0 \oplus e_1$. Binary operations evaluate under typical evaluation rules. We focus therefore on the substitution approach for each of the possible values:
\begin{itemize}
\item Values $v$ are already values and thus do not need to be substituted. 
\item Attributes $id[str]$ are substituted row-wise.
\end{itemize} 

\section{ONGOING IMPLEMENTATION}
\subsection{Naive Implementation}
To demonstrate the plausibility of a TQL-based data discovery system, we begin by outlining a naive implementation. This simple and intuitive approach, though unoptimized, is able to execute the core of TQL with full expressiveness. More specifically, this naive approach to TQL evaluates collections, operations and signatures fully through a breadth-first approach by procedurally computing table transformations and filters in direct correspondence with their type-collected relational algebra specifications.

\begin{itemize}
    \item \textbf{Signatures.} A collection signature restricts a collection to a subset which satisfies the associated predicate. In the naive approach, this is thus implemented by iterating over each element and filtering it by the predicate.
    \item \textbf{Operations.} A collection operation is formally defined as returning a collection containing the output of the underlying dataset function on all possible combinations of input collection datasets. As such, in the naive approach, each and all of these combinations are directly computed.
\end{itemize}

It should be apparent, by the construction of type-collected relational algebra, that evaluating individual primitive signatures and operations takes between linear and quadratic time (depending on the specific signature/operation) in this naive implementation. As such, the naive approach scales poorly with respect to the number of keywords in a query: nesting two parameter operations, for example, can produce queries which are evaluated in higher-order polynomial time. This is a major limitation, owing both to the enormity of modern databases and to the bottlenecks imposed on expressive power by slow evaluations of complex queries.

\subsection{Envisioned Architecture}

To create a truly practical data discovery system for TQL, we address these above identified efficiency concerns by propsing an alternative architecture that is depth-first and approximate. That is, we loosen the restriction that the resultant collection must contain all possible datasets satisfying the query, instead requiring only that a reasonable subset be produced. And in doing so, we enable the resultant collection to be constructed by repeated evaluation of the query statement on individually-selected datasets sampled from the input collections.

We claim that the consequent trade-off in correctness does not impair the effectiveness of the data discovery system. Indeed, to justify our assertion, it suffices to observe that most discovery system users seek only one (or at most, a few) "good enough" datasets to satisfy the needs of their task; as such, it is exceedingly unlikely that they would be interested in searching through the entire collection of all theoretically possible resultant datasets.

To further augment this approach, we leverage type inference (discussed in II.D, but omitted from the naive implementation) to apply restrictions on the input collections. In doing so, we reduce the number of possible samples for each input, thereby increasing the probability (and thus rate) at which valid combinations of input datasets are discovered.

\subsection{Ongoing Research into Type Inference}

Currently, the focus in TQL development has been on the formulation and formalization of its especially critical type inference system. We provide an overview of some results from ongoing work and also outline still-existent open questions.

To begin, recall that the objective of the type inference system is to determine the maximally-restrictive labels of input collections given some labels for some arbitrary collections in the query. To make such a determination, it is necessary to develop a methodology to infer signatures given mutual relations between collections.

Recently, we have finished formalizing the decidable rules of the type inference system. Taking inspiration from similar work in in type-checking information control for practical security languages \cite{Myers}, we have constructed a series of logical predicates which deterministically infer certain collection signatures from the collection to which they are subject. One straightforward such logical predicate is the inference that if collections $A$ and $B$ contain tables with columns $["ab"]$ and $["cd"]$ respectively, then the collection \texttt{JOIN} $A$ $B$ must contain tables with both column $["ab"]$ and column $["cd"]$.

In the majority of cases, however, the above formalization is insufficient insofar as many query contexts suggest many equally feasible inference possibilities. For example, the inverse of our previously discussed logical predicate does not necessarily hold: if \texttt{JOIN} $A$ $B$ contains tables with both column $["ab"]$ and column $["cd"]$, we do not necessarily know that the columns are distributed such that $A$ contains tables with columns $["ab"]$ and $B$ contains tables with columns $["cd"]$ since other valid distributions (e.g. $A$ containing both columns) are also possible. 

The development of an effective approach to tackle these inference ambiguities remains an open question. Presently, we are investigating the prospect of further extending the TQL type system (and associated inference rules) to be partially-probabilistic. Probabilistic type systems have been used to model the ambiguities of natural language semantics \cite{Cooper} and it is hoped that such a system, perhaps in combination with a Hidden Markov Model, might be sufficient to provide reasonably reliable inference for plausible query constructions.


\section{CASE STUDIES}


To demonstrate the practical advantages of TQL, we explore a select set of case studies through a few crucial data-related task archetypes:
\begin{itemize}
    \item \textbf{Constraint Search.} Tasks in which the goal is to search for all datasets from available data sources that satisfy some given property/relationship constraints, typically expressed as predicates.
    \item \textbf{Dataset Composition.} Tasks in which, given some already known/discovered datasets, the goal is to compose them to create a satisfactory resultant dataset.
    \item \textbf{Combined Queries.} Tasks in which the downstream transformation context of dataset composition is used to inform the appropriate datasets to be returned from a constraint search.
\end{itemize}

For each archetype, we present a couple of probable scenarios, and explore the extent to which TQL and the popular data discovery and retrieval languages of SRQL (from AURUM) \cite{Fernandez}, and SQL respectively are each able to solve the problems presented.

\subsection{Constraint Search}

For the first example of the constraint search archetype, we present the following scenario: Suppose that a user is looking for the economic data of various cities across the United States. They might choose to restrict their search space by limiting datasets to only those whose columns contain the keyword "gdp".

This scenario cannot be solved in SQL, barring some rather heterogeneous application of its language features, because SQL requires that all datasets used in a data retrieval query must be explicitly declared. 

By contrast, the scenario is easily and intuitively solved in both SRQL and TQL. In SRQL, the constraint desired in the scenario could be realized through the query:

\begin{flushleft}
    \texttt{results = schemaSearch("gdp")}
\end{flushleft}

And in TQL, the desired constraint could similarily be realized through the query:

\begin{flushleft}
    \texttt{ Q : \{COL*["gdp"]\} }
\end{flushleft}

Now, for the second example, we consider this more relational scenario: Suppose that a user is seeking to identify the extent to which some data is duplicated in their database. More specifically, suppose that they already have some dataset $A$ and that they seek to identify all datasets which are sufficiently semantically similar. 

Evidently, this task cannot be solved in SQL for the same reasons as our first example. In SRQL, however, such a scenario is explicitly solvable. Indeed, the structure of such a query is provided in the AURUM paper:

\begin{flushleft}
    \texttt{contentSim(table: str) = } \\
    \hspace*{5mm} \texttt{drs = columns(table)} \\
    \hspace*{5mm} \texttt{return jaccardContentSim(drs)} \\
    \hspace*{20mm} \texttt{OR cosineSim(drs)} \\
    \hspace*{20mm} \texttt{OR rangeSim(drs)} \\
    \texttt{results = contentSim("A")}
\end{flushleft}

In TQL, this relationship can also be expressed through a simple syntactic construction:

\begin{flushleft}
    \texttt{ Q : \{SIML[A]\} }
\end{flushleft}

\subsection{Dataset Composition}
For an example of the dataset composition archetype, we offer for consideration the scenario in which a user is already in possession of data about the GDP of different cities and separate data about the population of those cities. In this case, if the user sought to derive the additional insight by comparing GDP and population data, they might seek to compose the two datasets into a single query result by application of a join operation.

Quite evidently, such a scenario and any other scenarios within this archetype, cannot be accomplished in SRQL insofar as the language provides no expressions for useful data compositions.

By contrast, such a composition would be trivial to implement in either SQL or TQL. The following query from SQL would produce the desired result:

\begin{flushleft}
    \texttt{SELECT * FROM cities\_gdp} \\
    \texttt{INNER JOIN cities\_population} \\
    \texttt{ON cities\_gdp.nm = cities\_population.nm} \\
\end{flushleft}

And so would the following query from TQL:

\begin{flushleft}
    \texttt{JOIN[S["nm"] = T["nm"]]} \\
    \hspace*{5mm} \texttt{(S : \{SRC[cities\_gdp]\})} \\
    \hspace*{5mm} \texttt{(T : \{SRC[cities\_population]\})} 
\end{flushleft}

\subsection{Combined Queries}

Finally, let us now consider a scenario in which it would be necessary to conduct a constraint search and a dataset composition in tandem. 

Consider the case as follows: Suppose that the user suspects that a correlation exists between social media usage and obesity. The user is unsure if their exists a dataset containing information on both social media usage and obesity and thus they expect that they will need to join two datasets to produce a final dataset of interest.

In this scenario, neither SQL nor SRQL would be able to accomplish the task individually. SQL can join the datasets but it cannot find them, while SRQL can find the datasets but cannot join them. As such, significant manual input would be required from the user to find the datasets using SRQL, to join the datasets using SQL, and then to repeat the preceding process if the produced dataset was found to be inappropriate for the task.

By contrast, this tedious process could be accomplished by the user almost entirely automatically using TQL. Indeed, the user would need only choose the best dataset in the resultant collection produced by the following query:

\begin{flushleft}
    \texttt{(JOIN S T) : \{ } \\
    \hspace*{5mm} \texttt{COL*["obesity"] } \\
    \hspace*{5mm} \texttt{AND COL*["social media"]} \\
    \hspace*{5mm} \texttt{\}} \\
\end{flushleft}


\section{Conclusion}

In this paper, we provide a first introduction to TQL, a new query language that synthesizes search and transformation context to greatly increase expressive power. We also demonstrate, in the construction and optimization of TQL, the power of leveraging programming languages techniques to drive more flexible query language design. Further research is still needed to fully actualize TQL in a complete data discovery system. We hope that TQL will empower a more interdisciplinary and language-conscious approach to data discovery.



\bibliographystyle{abbrv}
\bibliography{sample-base}


\end{document}